# ACCELERATED LIFETIME TEST OF THE SRF DRESSED CAVITY/TUNER SYSTEM FOR LCLS II HE PROJECT*

Y. Pischalnikov[†], T. Arkan, C. Contreras-Martinez, B. Hartsell,

J. Kaluzny, R. Pilipenko, J.C. Yun, *Fermilab, Batavia, Illinois, USA*

W. Lahmadi, *Phytron, Inc.*

*Abstract*

The off-frequency detune method is being considered for application in the LCLS-II-HE superconducting linac to produce multi-energy electron beams for supporting multiple undulator lines simultaneously [1]. Design of the tuner has been changed to deliver roughly 3 times larger frequency tuning range. Working requirements for off-frequency operation (OFO) state that cavities be tuned at least twice a month. This specification requires the increase of the tuner longevity by 20 times compared with LCLS-II demands. Accelerated longevity tests of the LCLS-II HE dressed cavity with tuner were conducted at FNAL's HTS. Detail analysis of wearing and impacts on performances of the tuner's piezo and stepper motor actuators will be presented. Additionally, results of longevity testing of the dressed cavity bellow, when cooled down to 2 K and compressed by 2.6 mm for roughly 2000 cycles, will be presented.

## INTRODUCTION

LCLS II dressed cavity equipped with modified (LCLS II HE) tuner [2] installed in the FNAL HTS (Horizontal Test Stand). Cavity was cool down to 2K (and during initial part of test to 4K) and operated for almost 2 months.

Objective of this test was to demonstrate that stepper actuator and piezo actuators could withstand longevity specifications required for LCLS II HE OFO. Longevity of the Phytron stepper motor actuator [3], that deployed in the all the SRF cavity tuners of LCLS II Linac, reported in previous studies [4,5]. Specific of these studies to conduct test for significantly longer range and as close as possible to real LCLS II HE conditions. Additionally, this test was validated that more than 400 cycles of compression of bellow on 2.6mm will not damage bellow.

In the Table 1 specifications for tuner components (stepper motor actuator and piezo actuator) and dressed cavity for LCLS II HE OFO and LCLS II presented.

## LONGEVITY TEST AT HTS

Test has been run at HTS test 24 hours in day. LabView based program continuously run test and twice a day operator checked on progress. Test executed with following iterations: 1) read/record cavity frequency through NWA, 2) run stepper motor on 30kSteps, 3) read /record stepper motor temperature. Range of stepper motor (set by operator) was from 0 to +330kSteps that deliver cavity compression on 600kHz. One full cycle is 660kStep. Summary of the parameters for one tuner cycle presented on Table 2. LabView program has many interlocks: a) stepper motor overheating, b) frequency of cavity do not change after driver delivered on stepper motor pulses to move on 30kStep, c) failure of NWA to find the cavity frequency. As soon as any interlock will be triggered, LabView program will halt operation to prevent any damage to tuner/dressed cavity system.

Table 1. Specification for longevity of the stepper and piezo actuators for LCLS II & LCLS II HE

|  | LCSI II | LCLS II HE |
|---|---|---|
| Frequency tuning required for 95% of the cavities to bring to 1.3GHz after cooldown to T=2K, [ kHz] | 200 | 200 |
| Stroke/compression required to tune cavities to 1.3GHz after cooling down to T=2K, [mm] | 0.67 | 0.67 |
| Forces on the shaft/nut system to tune 95% of cavities to f=1.3GHz, [N] | 260 | 325 |
| Forces on the piezo actuator to tune 95% of cavities to f=1.3GHz, [kN] | 2.6 | 2.6 |
| Forces on the shaft/nut system to tune 95% of cavities to OFO f=1.3GHz-465kHz, [N] | N/A | 710 |
| Forces on the piezo actuator to tune 95% of cavities to OFO f=1.3GHz-465kHz, [kN] | N/A | 6 |
| Longevity of the actuator/Number of the motor steps to tune cavity from 1.3GHz to "safe" position before warm-up (twice a year) during 20 years, [MSteps] | 12 | 10 |
| Longevity of the actuator/Number of the motor steps to tune cavity from 1.3GHz to "1,3GHz-465kHz" and back 20 times a year during 20 years, [MSteps] | N/A | 210 |
| Longevity for 20 years operation, [Msteps] | 12 | **220** |
| Overall stroke of traveling nut on the shaft for 20 years of operation, [m] | 1.2 | 22 |
| Overall stroke/cavity compression for 20 years of operation, [m] | 0.03 | 0.69 |

Stepper motor located inside insulated vacuum environment and will overheating if run continuously. Working range for stepper motor was between T=40K to T=80K. This setting led to stepper work time ~ 110min and idle (cool-down time) ~62min. For one day (24hours) stepper tuner was able to run on ~ 10.3 MSteps that delivered 15.6 tuner's cycles. Summary of the HTS data accumulated for 2-month continuous test of tuner system presented on the Table 3.

All the data collected during 2-month HTS test (cavity frequency vs time) presented on the Figure 1. There are 2 distinct portions: operation at T=4K (at the beginning of test) and T=2K. Figure 1 demonstrated fact that stepper motor actuators (and whole tuner) do not experience any loss of performances after operating in cold/insulated vacuum environment for 414 MSteps that equivalent to 2 lifetimes of LCLS II HE OFO. Three plots/cycles (at start, middle and end of test) of cavity frequency retuning vs tuner stroke/stepper motor stroke coincided with very good accuracy.

Table 2. Forces on the tuner and actuators.

|  | dX, mm | df, kHz | Forces on tuner, kN | Forces on Stepper motor actuator shaft, N | Forces on the piezo actuator, kN |
|---|---|---|---|---|---|
| Cavity compression at start of cycle = minimim compression (from non-restrained position) | 0.6 | 200 | 2.4 | 150 | 1.2 |
| Cavity compression at the middle of cycle = maximim compression (from non-restrained position) | 2.6 | 800 | 10.4 | 650 | 5.2 |

Table 3. Summary of the HTS Accelerated Test of the longevity tuner/dressed cavity system

|  | Number of the cycles (660kSteps each cycle) | Accumulated number of steps by stepper motor, [MSteps] | Accumulated cavity re-tuning, [MHz] | Accumulated stroke/travel of the nut on the shaft, [m] | Accumulated stroke= compressions of the cavity, [m] |
|---|---|---|---|---|---|
| T=4K | 226 | 150 | 270 | 15 | 0.47 |
| T=2K | 400 | 264 | 475 | 26 | 0.8 |
| Total | 626 | 414 | 745 | 41 | 1.3 |

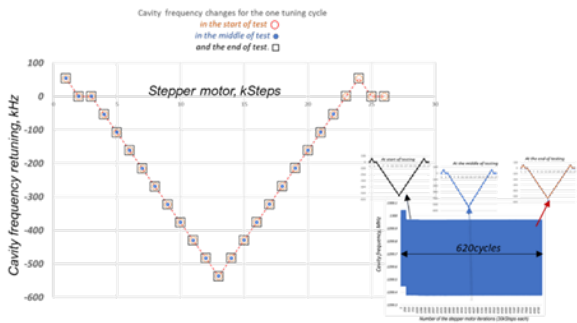

Figure 1. Cavity frequency changes for one full cycle (660kSteps) at the start, in the middle and end of test. No changes in cavity frequency tuning trend had been observed from start to the end of 626 cycles test (running tuner continuously for 2 month).

Someone could expect that operation of the stepper motor actuator for so many steps & overall distance of nut traveling on the shaft could lead to following failures (based on then existing experience with other actuator):

- ceasing of the traveling nut on the shaft,
- stripping thread of TECASIN portion of traveling nut,
- stripping dry lubrication from Titanium shaft,
- failure of the stepper motor (could be wearing of the dry lubrication on the bearing inside motor),
- failure of the planetary gear box (could be wearing of the dry lubrication on the bearing or other components of gear box),
- stepper missing steps that could be led to change of cavity frequency tuning for the same number of steps at start and end of test

Piezo-ceramic actuators P-844K075 [4], deployed in the tuner, had been subjected to 626 cycles with maximum forces up to ~6kN. Forces up to ~5.2kN derived from compressed cavity (Table 2) and 0.8kN is internal preload of stack inside capsulated actuator. We do not observe any piezo-actuators performances degradation after 2-month test.

## DRESSED CAITY'S BELLOW LONGEVITY STUDY

To demonstrate longevity of the dressed cavity bellow when compressed on 2.6mm (at T=2K) we need achieve ~2000. With tuner "as is" it will be required additional 4-5month of testing at HTS that could be difficult to accomplish for many reasons. We modify tuner to use steel cable that winding on the spool mounted on the Phytron shaft to compress cavity (figure 2). This modification allowed to decrease time for one cycle (with compression 2.6mm) in approximately 40 times compare with operation at HTS. Dressed cavity with "modified" tuner has been installed into VTS. There were no issues with overheating stepper motor when it is submerged into liquid helium. Operation of the tuner monitored by reading frequency of the cavity. During one day of operation at VTS (8-10 hours shift) modified tuner compressed cavity/below on ~200 cycles. In the end of each operational day cavity/bellow was warm up and inspection of the bellow (visual & pressure test) have been conducted. Dressed cavity/bellow system withstand without any failure of 1660 cycles (~8 lifetime) up to 2.6mm compression.

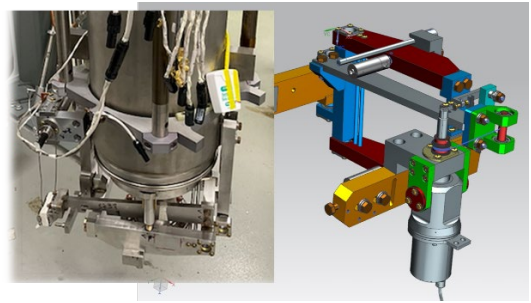

Fig 2: 3D model of the modified tuner. Picture of modified tuner mounted on the cavity before VTS test.

## INVESTIGATION OF THE IMPACT OF LONGEVITY TEST ON THE PHYTRON STEPPER ACTUATOR

Visual inspection of the stepper motor actuator has been performed after 2-month lifetime test at FNAL. During inspection (without dis-assembly of the actuator) main attention has been paid to spindle and traveling nut. Previous experience with longevity studies of different type of stepper actuator predicted that it could be most likely components of the actuators that could failed. We were not able to see any significant degradations (Figure 3).

As next step stepper actuator has been sent back to Phytron company for dis-assembly and detail investigation by experts from company that designed and manufactured actuator. Company conducted many different tests (visual, electrical, mechanical, etc.) before actuator was dis-assembled. After that actuator was dis-assembled and some actuator's components were cut. Detail investigation by optical microscope and with SEM (Scanning Electron Microscope) each component of the actuator (gear box, spindle system, stepper motor, bearing coated with dry lubricant) were performed at Phytron (figures 3,4,5).

Investigation demonstrated that some components show no wearing when other revealed some wear (figure 8). Ball bearing show signs of wear and failure of the adhesion of dry lubricant (DL-5 coating) (figure 5). During accelerated test at FNAL operation of the actuator exceed 40-50 times design specification. Phytron experts concluded that actuator has experienced a higher stress as it designed for and is still in good working condition.

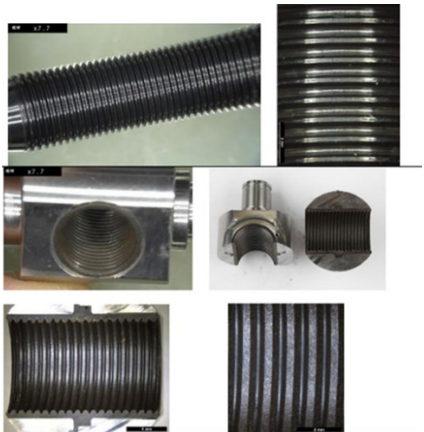

Figure 3 Pictures of Titanium shaft and TECASIN traveling nut. No visible wearing after 2-month longevity test.

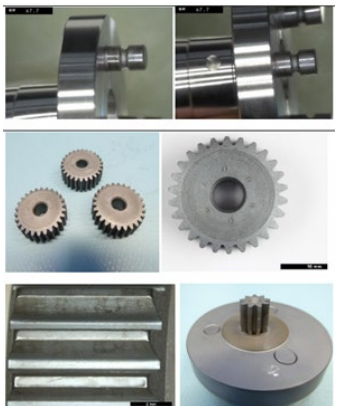

Figure 4. The second stage of gear box did not show significant damage.

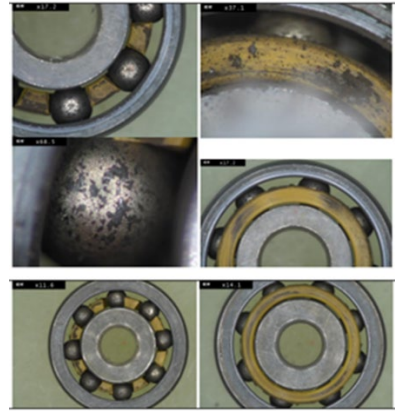

Figure 5. Ball bearing show signs of wear and failure of the adhesion of dry lubricant (DL-5 coating).

## CONCLUSION

Stepper motor actuator has been tested at cryogenic temperature and insulated vacuum during 2-month continuous operation. Accumulated number of steps was ~414 MSteps that is 2 times longer than estimated longevity for LCLS II HE (and in ~40 times longer that LCLS II longevity specifications). We do not observe any degradations of stepper actuator performances. Additional investigation of the stepper actuator by vendor (Phytron, Inc) show wearing of some components of actuator but not significant to impact overall performance of actuator. Dressed cavity/bellow system withstand without any failure of 1660 cycles up to 2.6mm compression that equivalent to ~8 lifetimes of LCLS II HE.